\documentstyle[12pt,a4wide,epsf]{article}

\begin{document}

\vspace{0.3cm}

\begin{center}
{\Large \bf FCNC transition $c\to u\gamma$ in $B_c\to B_u^*\gamma$ decay }

\vspace{0.5cm}

{S. Fajfer$^{a,b}$\footnote{At Heavy Flavours 8 presented by S. Fajfer; e-mail: svjetlana.fajfer@ijs.si} and S. Prelov\v sek$^{a}$}

\vspace{0.15cm}

{\it a) J. Stefan Institute, Jamova 39, 1001 Ljubljana, Slovenia}

{\it b) Department of Physics, University of Ljubljana, 1000 Ljubljana, 
Slovenia} 

\vspace{0.5cm}

{P. Singer}

\vspace{0.15cm}

{\it  Department of Physics, Technion - Israel Institute  of Technology, 
Haifa 32000, Israel}
\end{center}

\vspace{0.8cm}

\centerline{\large \bf Abstract}

\vspace{0.3cm}

We propose the $B_c\to B_u^*\gamma$ decay as the most suitable  probe 
for the 
flavour changing neutral transition $c\to u\gamma$.  We estimate the short and 
long distance contributions to this decay within the standard model and we find 
them to be comparable; this is in contrast to radiative decays of $D$ mesons, 
that are completely dominated by the long distance contributions. Since the 
$c\to u\gamma$ transition is very sensitive to the physics beyond the standard 
model, the standard model prediction $Br(B_c\to B_u^*\gamma)\sim 10^{-8}$ 
obtained  here opens a new window for future experiments.    
The detection of $B_c\to B_u^*\gamma$ decay at branching ratio well above 
$10^{-8}$ would signal new physics.

\section{Introduction}

Flavour changing neutral current (FCNC) transitions occur in the standard model 
only at the loop level. Hence, they are very rare in the standard model and they 
present a suitable probe for new physics. The FCNC transitions in the down-quark 
sector are relatively frequent due to the large mass of the top quark  running 
in the loop and the transition $b\to s$  has indeed been observed \cite{CLEO}. 
The FCNC transitions in the up-quark sector are especially rare in the standard 
model due to the small masses of the intermediate down-like quarks that run in 
the loop. For these transitions, the standard model represents a small 
background for the possible contributions arising from some new physics. At 
present, only upper experimental limits on the FCNC transitions in the up-quark 
sector are available \cite{PDG}.

We study the transition $c\to u\gamma$, which is the most 
probable FCNC transition in the up-quark sector within the standard model. To 
probe the $c\to u\gamma$ transition we propose the radiative beauty-conserving 
decay $B_c\to B_u^*\gamma$ \cite{FPS}; the  $B_c$ meson has been detected  recently at 
Fermilab \cite{CDF}. We estimate the short distance (SD) and long distance (LD) 
contributions to $B_c\to B_u^*\gamma$ decay \cite{FPS} within the standard model. The most serious among long distance contributions  is illustrated at the quark level in Fig. 1. It  is proportional to the small CKM factor $V_{cb}^*V_{ub}$ and it is therefore relatively small. The short and long distance contributions to $B_c\to B_u^*\gamma$ are found to be comparable \cite{FPS}, which allows us in principle to probe $c\to u\gamma$ 
transition in this decay. 
This is in contrast to the case of $D$ meson decays, where the $b$ quark is replaced by $d$ or $s$ quark in Fig. 1 and the corresponding long distance contribution is proportional to the relatively big CKM factors $V_{cd}^*V_{ud}$ or $V_{cs}^*V_{us}$. 
As a consequence, the radiative $D$ meson decays are  completely dominated by the LD contributions [5-9] 
and it is impossible to extract the short distance  $c\to u\gamma$ contribution from the experiment. 

\begin{figure}
 \begin{center}
 \leavevmode\epsfbox{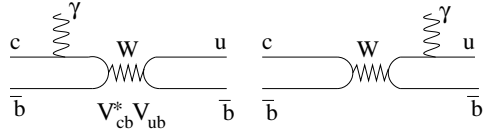} 
 \caption{The most serious among the long distance contributions (called the pole contribution) to $B_c\to B_u^*\gamma$ decay at the quark level. The photon can be emitted from any of the quark lines. } 
 \label{fig1}
 \end{center}
  \end{figure}

\section{ The short distance contribution}

The SD contribution in $B_c\to B_u^*\gamma$ decay is driven by FCNC $c\to 
u\gamma$ transition and $\bar b$  is a spectator. The $c\to u\gamma$ transition 
is strongly GIM suppressed at one-loop, QCD logarithms enhance the amplitude by 
two orders of magnitude \cite{GHMW}, while the complete 2-loop QCD corrections 
further increase the amplitude by two orders of magnitude \cite{GHMW}. The 
Lagrangian that induces the $c\to u\gamma$ transition is given by
\begin{eqnarray}
{\cal L}_{SD}^{c\to u\gamma}&=&-{G_F\over \sqrt{2}}{e\over 
4\pi^2}V_{cs}V_{us}^*~c_7^{c\to u\gamma}(\mu)\nonumber\\
&\times&\bar 
u\sigma^{\mu\nu}[m_c{1+\gamma_5\over 2}+m_u{1-\gamma_5\over 2}]c~F_{\mu\nu}~.\nonumber
\end{eqnarray}
The appropriate renormalization scale $\mu$ for $c_7^{c\to u\gamma}$ in $B_c\to B_u^*\gamma$ 
decay is $\mu=m_c$ (and not $\mu=m_b$), since $\bar b$ is merely a spectator in 
the SD process. The 2-loop QCD calculation was performed in  \cite{GHMW}, giving 
$c^{c\to u\gamma}_{7}(m_c)=-0.0068-0.020i$. 

The corresponding amplitude for $B_c\!\!\to\!\! B_u^*\gamma(q,\!\epsilon)$ decay is 
proportional to $\epsilon_{\mu}^*q_{\nu}\langle B_u^*|\bar 
u\sigma^{\mu\nu}(1\pm\gamma_5)c|B_c\rangle$ taken at $q^2=0$, which can be 
expressed in terms of the form factors $F_1(0)$ and $F_2(0)$ \cite{Soares}:
\begin{eqnarray}
\label{form1}
&\epsilon_{\mu}^*&\!\!\!\langle B_u^*(p^{\prime},\epsilon^{\prime})|\bar 
ui\sigma^{\mu\nu}q_{\nu}c|B_c(p)\rangle_{q^2=0}=\nonumber\\
&=&i\epsilon^{\mu\alpha\beta\gamma}\epsilon_{\mu}^*
\epsilon_{\alpha}^{*\prime}p_{\beta}^{\prime}p_{\gamma}F_1(0)~
,\nonumber\\
&\epsilon_{\mu}^*&\!\!\!\langle B_u^*(p^{\prime},\epsilon^{\prime})|\bar 
ui\sigma^{\mu\nu}q_{\nu}\gamma_5c|B_c(p)\rangle_{q^2=0}=\nonumber\\
&=&[(m_{B_c}^2-m_{B_u^*}^
2)\epsilon^*\cdot\epsilon^{*\prime}-2(\epsilon^{*\prime}\cdot q)(p\cdot 
\epsilon^{*})]F_2(0).\nonumber\\
~
\end{eqnarray}
 The form factors defined will be calculated using the 
ISGW model \cite{ISGW}. \\

\section{The long distance contributions}

The long distance contributions are calculated using the nonleptonic weak 
Lagrangian \cite{FS} 
\begin{eqnarray}
\label{LD}
&{\cal L}^{eff}&\!\!\!=\!-{G_F\over \sqrt{2}}V_{uq_i}\!V^*_{cq_j}
[a_1\bar u\gamma^{\mu}(1\!-\!\gamma_5) 
q_i\bar q_j\gamma_{\mu}(1\!-\!\gamma_5)c\nonumber\\
&+&\!\!\!a_2\bar u \gamma^{\mu}(1\!-\!\gamma_5)c)\bar q_j\gamma_{\mu}(1\!-\!\gamma_5)q_i]~,
\end{eqnarray}
where  $q_i$, 
$q_j$ are the down quarks $d$, $s$, $b$ and $a_1$, $a_2$ include the QCD 
corrections \cite{BSW}.

Quite generally, the LD contributions to $B_c\to B_u^*\gamma$ decay can be 
separated into two classes \cite{FPS} related to the two terms of  (\ref{LD}), as performed 
previously \cite{BGHP} for $D\to V\gamma$ decays. The class (I), called also the 
vector meson dominance (VMD) contribution, is related to the $a_2$ term (\ref{LD}) and  corresponds to 
the processes $c\to u\bar q_iq_i$ followed by $\bar q_iq_i\to \gamma$, while 
$\bar b$ is the spectator in $B_c\to B_u^*\gamma$ decay. At the hadron level the 
$\bar q_iq_i\to \gamma$ transition is expressed using the vector meson dominance 
(VMD) and the corresponding diagram is depicted in Fig. (2a). The class (II), 
called also the {\it pole} contribution, is the most serious long distance contribution and it is presented at the quark level in Fig. 1. It is related to the $a_1$ term (\ref{LD}) 
and  corresponds to the process $c\bar b\to u\bar b$ with the photon attached to 
incoming or outgoing quark lines. Selecting the lowest contributing states, the 
pole contributions are depicted in Fig. (2b) at the hadron level. 

\begin{figure}
 \begin{center}
 \leavevmode\epsfbox{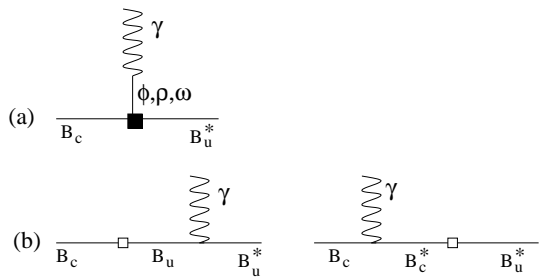} 
 \caption{ Long distance contributions in $B_c\to B_u^*\gamma$ decay. a) VMD 
contribution; the black box denotes the action of the Lagrangian (\ref{VMD}). 
b) {\it pole} contribution presented in Fig. 1 at the quark level; the white box denotes the action of the Lagrangian 
(\ref{pole}).} 
 \label{fig2}
 \end{center}
  \end{figure}

We turn now to the estimation of these two classes of contributions and we start 
with the VMD contribution (class (I)) represented by Fig. (2a). The underlying 
quark processes are $c\to u\bar ss(\bar dd)$ with $\bar ss,~\bar dd$ hadronizing 
into vector mesons $\phi$, $\rho$, $\omega$ which then turn to a photon, while 
$\bar b$ remains a spectator. We neglect the contribution of $\bar bb\to \gamma$ 
in view of the large mass of $\Upsilon$. The relevant part of the Lagrangian, 
after using the relations among CKM matrix elements, is
\begin{eqnarray}
\label{L1}
{\cal L}_{(I)}^{eff}&=&-{G_F\over \sqrt{2}}a_2(\mu)V_{cs}V_{us}^*~\bar 
u\gamma^{\mu}(1-\gamma_5)c\nonumber\\
&\times&\!\![\bar s\gamma_{\mu}(1-\gamma_5)s-\bar 
d\gamma_{\mu}(1-\gamma_5)d].
\end{eqnarray}
The appropriate scale for $a_2(\mu)$ in $B_c\to B_u^*\gamma$ decay is $\mu=m_c$, 
since $\bar b$ is again merely a spectator in VMD contribution. 
Thus, we may use $a_2(m_c)=-0.5$, as obtained in the successful phenomenological 
fit to $D$ meson decays \cite{BSW}. Defining $\langle 
V(q,\epsilon)|V_{\mu}|0\rangle=g_V(q^2)\epsilon_{\mu}^*$ and using the 
factorization approximation, the effective Lagrangian that induces the VMD 
contribution is given by 
\begin{eqnarray}
\label{VMD}
{\cal L}_{VMD}^{c\to u\gamma(\epsilon)}&=&-{G_Fe\over 
\sqrt{2}}a_2(m_c)V_{cs}V_{us}^*C_{VMD}^{'}\nonumber\\
&\times&\bar u\gamma^{\mu}(1-\gamma_5)c~\epsilon_{\mu}^*~,
\end{eqnarray}  
where
\begin{eqnarray} 
\label{cvmd}
C_{VMD}^{'}&=&{g_{\rho}^2(0)\over  2m_{\rho}^2}-{g_{\omega}^2(0)\over  
6m_{\omega}^2}-{g_{\phi}^2(0)\over  3m_{\phi}^2}\nonumber\\
&=&(-1.2\pm 1.2)\cdot 
10^{-3}~GeV^2~
\end{eqnarray}
is obtained by assuming  $g_{V}(m_{V})=g_{V}(0)$, with the mean value and the 
error in (\ref{cvmd}) calculated from the experimental data on $\Gamma(V\to 
e^+e^-)$ \cite{PDG}. Note here the remarkable GIM cancellation carried over to 
the hadronic level. 

Lagrangian (\ref{VMD}) implies that the  VMD amplitude for $B_c\to 
B_u^*\gamma(q,\epsilon)$ is proportional to  $\epsilon_{\mu}^*\langle B_u^*|\bar 
u\gamma^{\mu}(1-\gamma_5)c|B_c\rangle$ taken at $q^2=0$. For the hadronic matrix 
elements, one defines appropriate form factors for the vector and axial 
transitions as follows \cite{Soares}:
\begin{eqnarray} 
\label{form2}
\langle B_u^*(p^{\prime},\epsilon^{\prime})\!\!\!\!&|&\!\!\!\! \bar 
u\gamma^{\mu}(1-\gamma_5)c|B_c(p)\rangle=\\
&-&{2i\over m_{B_c}+m_{B_u^*}}\epsilon^{\mu\alpha\beta\gamma}\epsilon_{\alpha}^{*\prime}p_{\
beta}^{\prime}p_{\gamma}V(q^2)\nonumber\\
&+&(m_{B_c}+m_{B_u^*})\epsilon^{\mu 
*\prime}A_1(q^2)\nonumber\\
&-&{\epsilon^{*\prime}\cdot q\over 
m_{B_c}+m_{B_u^*}}(p+p^{\prime})^{\mu}A_2(q^2)\nonumber\\
&-&2m_{B_u^*}{\epsilon^{*\prime}\cdot q\over q^2}q^{\mu}[A_3(q^2)-A_0(q^2)]~.\nonumber
\end{eqnarray}
The requirements of the finite matrix elements at $q^2=0$ \cite{BSW} and of 
gauge invariance lead to the relations among the various form factors 
\cite{FPS1}, which imply $A_0(0)=A_3(0)=0$ and 
$A_2(0)=[(m_{B_c}+m_{B_u^*})/(m_{B_c}-m_{B_u^*})]A_1(0)$. The same relations are 
obtained by using the prescription that the photon couples only to the 
transverse polarization of the current \cite{FPS1,GP}. Accordingly, the VMD 
amplitude will  be expressed in terms of two form factors only, $V(0)$ and 
$A_1(0)$. 

At this point, we remark that the form factors $F_1$, $F_2$, $V$ and $A_1$, 
needed for the SD and VMD amplitudes cannot be safely related using the 
Isgur-Wise relations \cite{IW}, since the masses of $b$ and $c$ quarks composing 
$B_c$ meson do not permit the $\bar b$ quark to be at rest. Therefore we shall 
determine the corresponding form factors at $q^2=0$ independently, using the 
ISGW model \cite{ISGW}.

\vspace{0.2cm}

We now turn to the discussion of the LD contributions of class (II), the {\it 
pole} contribution, where the quark process $c \bar b \to u\bar b$ is driven by
\begin{equation}
\label{pole}
{\cal L}_{(II)}^{eff}=-{G_F\over \sqrt{2}}a_1(\mu)V_{cb}V_{ub}^*~\bar 
u\gamma^{\mu}(1-\gamma_5)b~\bar b\gamma_{\mu}(1-\gamma_5)c~
\end{equation}
and the photon line is attached to any of four quark lines.  In terms of 
hadronic degrees of freedom this diagram is given in Fig. (2b), where the white 
box represents the action of the Lagrangian (\ref{pole}) (we have neglected the 
contribution of the scalar and axial poles).  
Considering the scale for $a_1(\mu)$ in $c\bar b\to u\bar b$, it is difficult to 
decide between $\mu=m_c$ or $\mu=m_b$, since $\bar b$ is not spectator in the 
pole contribution. As the difference between  $a_1(m_c)=1.2$ and $a_1(m_b)=1.1$ 
\cite{BSW} and is not essential, we take $a_1(m_b)=1.1$. 
Note that the pole contribution is relatively small due to the factor 
$V_{cb}V_{ub}^*$ in (\ref{pole}). In $D$ meson decays, the corresponding factor 
$V_{cs}V_{us}^*$ is much bigger, which makes the pole contribution dominant  
over the SD and VMD ones \cite{BGHP,FS,FPS1}. {\bf Different CKM factors  
in the pole contribution of $B_c$ and $D$ decays are essential in 
establishing 
the $B_c\to B_u^*\gamma$ decay as more suitable  for  the investigation of $c\to 
u\gamma$ than the $D$ decays.}

To evaluate the amplitude for the pole diagrams given in Fig. (2b) we define    
\begin{eqnarray}
\label{pole1}
\langle 0|A_{\mu}|P\rangle &=&f_Pp_{\mu}\\
\langle V|V_{\mu}|0\rangle &=&g_V\epsilon_{\mu}^*\nonumber\\
{\cal A}(P(p)\to V(p^{\prime},\epsilon^{\prime})\gamma(\epsilon 
))&=&\mu_Pe\epsilon^{\mu\nu\alpha\beta}\epsilon_{\mu}^*\epsilon_{\nu}^{*\prime}p
_{\alpha}p_{\beta}^{\prime}~,\nonumber
\end{eqnarray}
where $\mu_{B_c}$, $\mu_{B_u}$, $f_{B_c}$, $f_{B_u}$, $g_{B_c^*}$ and 
$g_{B_u^*}$ will be determined using ISGW model.\\

\section{The amplitude}

Using the above Lagrangians and form factor decomposition of Eqs. (\ref{form1}), 
(\ref{form2}), (\ref{pole1}), the final amplitude for $B_c\to B_u^*\gamma$ 
containing SD and LD contributions can be expressed as
\begin{eqnarray}
\label{amp}
 A\!\!\!&(&\!\!\!B_c(p)\to B_u^*(p^{\prime},\epsilon^{\prime})\gamma(q,\epsilon 
))=\nonumber\\
&+&i\epsilon_{\mu}^{*\prime}\epsilon_{\nu}^*[A_{PV}(p^{\mu}p^{\nu}-g^{\mu\nu}p\dot q)\nonumber\\
&+&iA_{PC}\epsilon^{\mu\nu\alpha\beta}p^{\prime}_{\alpha}p_{\beta}]~,
\end{eqnarray}
where
\begin{eqnarray}
\label{e1}
&A_{PV}&\!\!=\!-{G_F\over \sqrt{2}} e\biggl(V_{cs}V_{ud}^*\biggl[{c_7^{c\to 
u\gamma}(m_c)\over 2\pi^2}(m_c-m_u)F_2(0)\nonumber\\
\!\!&+&\!\!\!\!\! 2a_2(m_c)C_{VMD}^{\prime}{A_1(0)\over 
m_{B_c}-m_{B_u^*}}\biggr]\biggr)~,\nonumber\\
&A_{PC}&\!\!=\!-{G_F\over \sqrt{2}} e\biggl(V_{cs}V_{ud}^*\biggl[{c_7^{c\to 
u\gamma}(m_c)\over 4\pi^2}(m_c+m_u)F_1(0)\nonumber\\
\!\!&+&\!\!\!\!\! 2a_2(m_c)C_{VMD}^{\prime}{V(0)\over 
m_{B_c}+m_{B_u^*}}\biggr]\nonumber\\
\!\!&+&\!\!\!\!\!\!V_{cb}V_{ub}^*a_1\biggl[{\mu_{B_c}g_{B_c^*}g_{B_u^*}\over 
m_{B_c^*}^2-m_{B_u^*}^2}\!+\!{\mu_{B_u}m_{B_c}^2f_{B_c}f_{B_u}\over 
m_{B_c}^2-m_{B_u}^2}\biggr]\nonumber
\biggr).
\end{eqnarray}
The first term in Eqs. (\ref{e1}) comes from SD contribution, the second term 
from VMD contribution and the third term from the {\it pole} contribution. The 
decay width is then given by
\begin{equation}
\label{gamma}
\Gamma={1\over 4\pi}\biggl({m_{B_c}^2-m_{B_u^*}^2\over 
2m_{B_c}}\biggr)^3(|A_{PV}|^2+|A_{PC}|^2)~.
\end{equation}

\section{The model}

To account for the nonperturbative dynamics wit\-hin the mesons we use the 
nonrelativistic constituent ISGW quark model \cite{ISGW}. This model is 
considered  to be reliable for a state composed of two heavy quarks, which makes 
it suitable for treating $B_c$; in addition the velocity of $B_u^*$ in the rest 
frame of $B_c$ is to a fair approximation nonrelativistic. In the ISGW model the 
constituent quarks of mass $M$ move under the influence of the effective 
potential $V(r)=-4\alpha_s/(3r)+c+br$, $c=-0.81~GeV$, $b=0.18~GeV^2$ 
\cite{ISGW2}. Instead of the accurate solutions of the Schrodinger equation, the 
variational solutions 
$$\psi(\vec r)=\pi^{-{3\over 4}}\beta^{{3\over 2}}e^{-{\beta^2r^2\over 
2}}~~~{\rm or}~~~ \psi(\vec k)=\pi^{-{3\over 4}}\beta^{-{3\over 2}}e^{-{k^2\over 
2\beta^2}}$$
for $S$ state are used, where $\beta$ is employed as the variational parameter. The meson 
state composed of constituent quarks $q_1$ and $\bar q_2$ is given by   
\begin{eqnarray}
&|&\!\!\!\!M(p)\rangle=\sum_{C,s1,s2}{1\over \sqrt{3}}\sqrt{{2E\over (2\pi)^3}}\int d\vec 
k\psi(\vec k)\sqrt{{M_1\over E_1}}\sqrt{{M_2\over 
E_2}}\nonumber\\
&\times& \!\!\!f_{s2,s1}\delta(p-p_1-p_2)b_1^\dagger(\vec p_1,s_1,C)d_2^\dagger(\vec 
p_2,s_2,\bar C)|0\rangle~,\nonumber
\end{eqnarray} 
where $\vec k$ is the momentum of the constituents in the meson rest frame, $C$ 
denotes the colour, while 
$f_{s2,s1}=(\bar\uparrow\downarrow+\bar\downarrow\uparrow)/\sqrt{2}$ for 
pseudoscalar and 
$f_{s2,s1}=(\bar\uparrow\downarrow-\bar\downarrow\uparrow)/\sqrt{2},\bar\uparrow
\uparrow,\bar\downarrow\downarrow$ for vector mesons. Using the normalization of 
the spinors as in \cite{itzykson}, we obtain in the nonrelativistic limit 
\begin{eqnarray}
V(q^2)&=&{m_{B_c}+m_{B_u^*}\over2}F_3(q^2)\nonumber\\
&\times&\biggl[{1\over 
M_u}-{M_b(M_c-M_u)\beta_{B_c}^2\over 
M_cM_um_{B_u^*}(\beta_{B_c}^2+\beta_{B_u^*}^2)}\biggr]~,\nonumber\\
A_1(q^2)&=&F_2(q^2)={2m_{B_c}\over m_{B_c}+m_{B_u^*}}F_3(q^2)~,\nonumber\\
F_1(q^2)&=&2F_3(q^2)\biggl[1+(m_{B_c}-m_{B_u^*})\nonumber\\
&\times&\biggl({1\over 
2M_u}-{M_b(M_c+M_u)\beta_{B_c}^2\over 
2M_cM_um_{B_u^*}(\beta_{B_c}^2+\beta_{B_u^*}^2)}\biggr)\biggr]~,\nonumber\\
\mu_{B_c}&=&\sqrt{{m_{B_c^*}\over 
m_{B_c}}}\biggl({2\beta_{B_c}\beta_{B_c^*}\over 
\beta_{B_c}^2+\beta_{B_c^*}^2}\biggr)^{3\over 2}\biggl[{2\over 3M_c}-{1\over 
3M_b}\biggr]~,\nonumber\\
f_{B_c}&=&{2\sqrt{3}\beta_{B_c}^{{3\over 2}}\over \pi^{{3\over 
4}}\sqrt{m_{B_c}}}~,\nonumber\\
g_{B_c^*}&=&m_{B_c^*}{2\sqrt{3}\beta_{B_c^*}^{{3\over 2}}\over \pi^{{3\over 
4}}\sqrt{m_{B_c^*}}}
\end{eqnarray}
and analogously for $\mu_{B_u}$, $f_{B_u}$ and $g_{B_u^*}$. Here 
\begin{eqnarray}
F_3(q^2)&=&\sqrt{m_{B_u^*}\over m_{B_c}}\biggl({2\beta_{B_c}\beta_{B_u^*}\over 
\beta_{B_c}^2+\beta_{B_u^*}^2}\biggr)^{3/2}\nonumber\\
&\times&\exp\biggl(-{M_b^2\over 
2m_{B_c}m_{B_u^*}} {[(m_{B_c}-m_{B_u^*})^2-q^2]\over 
\kappa^2(\beta_{B_c}^2+\beta_{B_u^*}^2)}\biggr)~,\nonumber
\end{eqnarray}
where $\kappa=0.7$ \cite{ISGW}. The results for $V(q^2)$ and $A_1(q^2)$ 
reproduce the results of \cite{ISGW}, while $F_1(q^2)$ and $F_2(q^2)$ represent, 
to our knowledge, the new results within ISGW model. Using parameters $\beta$ 
\cite{ISGW2} and meson masses given in Table 1  and the constituent quark masses 
$M_u=0.33~GeV,~~M_c=1.82~GeV~~{\rm and}~~M_b=5.2~GeV$ \cite{ISGW2} we get
$$f_{B_u}\!\!=\!0.18~GeV,~g_{B_u^*}\!=\!0.86~GeV^2,~\mu_{B_u}\!=\!1.81~GeV^{-1}$$
$$f_{B_c}\!\!=\!0.51~GeV,~g_{B_c^*}\!=\!2.41~GeV^2,~\mu_{B_c}\!=\!0.28~GeV^{-1}$$
while the form factors evaluated at $q^2=0$ are given in Table 2. \\

\begin{table}[h]
\begin{center}
\begin{tabular}{|c||c|c|c|c|}
\hline
& $B_c$&$B_c^*$&$B_u$&$B_u^*$\\
\hline
$m$&6.40 \cite{CDF}&6.42 \cite{ISGW2}&5.28 \cite{PDG}&5.325 \cite{PDG}\\
\hline
$\beta$&0.92&0.75&0.43&0.40\\
\hline
\end{tabular}
\caption{Parameters $\beta$ (taken from \cite{ISGW2})  and masses of 
pseudoscalar and vector mesons in GeV.  }
\end{center}
\end{table}

\begin{table}[h]
\begin{center}
\begin{tabular}{|c|c|c|c|}
\hline
$A_1(0)$&$V(0)$&$F_1(0)$&$F_2(0)$\\
\hline
0.24&1.3&0.48&0.24\\
\hline
\end{tabular}
\caption{The $B_c\to B_u^*$ form factors at $q^2=0$ calculated using ISGW model \cite{ISGW}.  }
\end{center}
\end{table}

\section{ The results }

We use the central value of the current quark masses $m_u=0.0035~GeV$, 
$m_c=1.25~GeV$ from \cite{PDG} and $V_{cb}=0.04$, $V_{ub}=0.0035$. The SD, VMD 
and {\it pole} contributions to amplitudes $A_{PC}$ and $A_{PV}$ needed to 
compute the amplitude (\ref{amp}) and the decay rate (\ref{gamma}) are given in 
Table 3, where the error is due only to the uncertainty in parameter 
$C_{VMD}^{\prime}$ (\ref{cvmd}). In Table 4 we present the total branching ratio 
and separately also the SD and LD part of the branching ratios for $B_c\to 
B_u^*\gamma$ decay, where we have taken $\tau(B_c)=0.46{+0.18\atop 
-0.16}\pm0.03~ps$ as measured by CDF Collaboration recently \cite{CDF}. Note 
that SD and LD contributions  give branching ratios of comparable size $\sim 
10^{-8}$, which in principle allows to probe the $c\to u\gamma$ transition in 
$B_c\to B_u^*\gamma$ decay. Experimental detection of $B_c\to B_u^*\gamma$ decay 
at the branching ratio well above $10^{-8}$ would clearly indicate a signal for 
new physics.  The measurement of this decay would probe different scenarios of  
physics beyond the standard model: the non-minimal supersymmetric model 
\cite{BGM} and the standard model with four generations \cite{BHLP}, for 
example, predict  $Br(c\to u\gamma)$ up to $10^{-5}$, which would enhance 
$Br(B_c\to B_u^*\gamma)$ up to $10^{-6}$.   The branching ratios for $D$ meson decays with flavour content $c\bar q\to 
u\bar q\gamma$, on the other hand, are of order $10^{-6}$ 
even within the standard model \cite{FS,FPS1,GHMW}: they are driven mainly by 
the long distance pole contributions analogous to those in Fig. 1, which overshadow the  $c\to u\gamma$ 
transition (predicted at the branching ratio $\sim 10^{-9}$ in the standard 
model) and possible signals of new physics.\\

\begin{table}[h]
\begin{center}
\begin{tabular}{|c|c|c|}
\hline
 $A_{PV}^{SD}$ & $A_{PV}^{VMD}$ & $A_{PV}^{pole}$  \\
\hline
$5.7+17~i$&$-14\pm 14$&$0$\\
\hline
\hline
$A_{PC}^{SD}$ & $A_{PC}^{VMD}$ & $A_{PC}^{pole}$ \\
\hline
$5.7+17~i$&$-7.3\pm 7.3$&$-21$\\
\hline
\end{tabular}
\caption{ The parity conserving (PC) and parity violating (PV) amplitudes (\ref{amp}) for $B_c\to B_u^*\gamma$ decay. The short distance (SD), vector meson dominance (VMD) and {\it 
pole} contributions as predicted by ISGW model are given separately in units of $10^{-11}~GeV^{-1}$. The error-bars  are due to the uncertainty in 
$C_{VMD}^{'}=(1.2\pm 1.2)~10^{-3}~GeV^2$ (\ref{cvmd}).}
\end{center}
\end{table}

\begin{table}[h]
\begin{center}
\begin{tabular}{|c|c|c|c|}
\hline
 $Br^{SD}$ & $Br^{LD}$ & $Br^{tot}$ \\
\hline
 $4.7\cdot 10^{-9}$ & $(7.5 {+7.7\atop -4.3})\cdot 10^{-9}$ 
& $(8.5 {+5.8 \atop -2.5})\cdot 10^{-9}$\\
\hline
\end{tabular}
\caption{The total branching ratio for $B_c\to B_u^*\gamma$ decay and its short distance (SD) and long distance (LD) parts as predicted by ISGW model. The 
error-bars are due to the uncertainty in 
$C_{VMD}^{'}=(1.2\pm 1.2)~10^{-3}~GeV^2$ (\ref{cvmd}).}
\end{center}
\end{table} 

\section{Summary} 

\vspace{0.3cm}

The long distance contribution in $B_c\to B_u^*\gamma$ decay is relatively small and this decay is proposed as the most suitable decay to probe the flavour changing neutral transition $c\to u\gamma$. 
The short distance part (driven by $c\to u\gamma$) and the long distance part of 
the branching ratio for $B_c\to B_u^*\gamma$ decay, presented in Table 4, are of comparable 
size. They are both of order $10^{-8}$ and the short distance contribution can in principle be  disentangled in this channel.  Since $c\to 
u\gamma$ transition is sensitive to the physics beyond the standard model, 
it would be very desirable to compare the standard model prediction of 
$Br(B_c\to B_u^*\gamma)=(8.5 {+5.8 \atop -2.5})\cdot 10^{-9}$ presented here to 
the experimental data in the future.    
The detection of $B_c\to B_u^*\gamma$ decay at a branching ratio well above 
$10^{-8}$ would signal new physics. In comparison to $B_c\to B_u^*\gamma$ decay, 
the  $D$ meson decays are far less suitable for probing $c\to u\gamma$ 
transition, since they are almost completely dominated by the long distance 
effects.

Finally, we wish to stress that $B_c\to B_u^*\gamma$ 
is characterized by a very clear
signature: their detection requires the observation of a $B_u/B_d$ decay in
coincidence with two photons. The $B_c\to B_u^*\gamma$  transition involves the 
emission of a high energy ($985~ MeV$) and of a low energy ($45~ MeV$) photon in 
the respective centers of mass of $B_c$, $B_u^*$.


\begin{thebibliography}{10} 
\bibitem{CLEO} R. Ammar {\it et al.} (CLEO Collab.), Phys. Rev. Lett. {\bf 71}, 
674 (1993); M. S. Alam {\it et al.} (CLEO Collab.), Phys. Rev. Lett. {\bf 74}, 
2885 (1995).   
\bibitem{PDG} Particle Data Group, Eur. Phys. J. C {\bf 3}, 1 (1998).
\bibitem{FPS} S. Fajfer, S. Prelov\v sek and P. Singer, Phys. Rev. D {\bf 59}, 
114003 
(1999).
\bibitem{CDF} CDF Collaboration, Phys. Rev. Lett. {\bf 81}, 2432-2437 (1998).
\bibitem{BGHP} G. Burdman, E. Golowich, J.L. Hewett and S. Pakvasa, Phys. Rev. D 
{\bf 52}, 6383 (1995).
\bibitem{FS} S. Fajfer and P. Singer, Phys. Rev. D {\bf 56}, 4302 (1997).
\bibitem{FPS1} S. Fajfer, S. Prelov\v sek and P. Singer, Eur. Phys. J. C {\bf 
6}, 471 (1999).  
\bibitem{FPS2} S. Fajfer, S. Prelov\v sek and P. Singer, Phys. Rev. D {\bf 58} 
094038 (1998).
\bibitem{GHMW} G. Greub, T. Hurth, M. Misiak and D. Wyler, Phys. Lett.  B {\bf 
382}, 415 (1996).
\bibitem{Soares} J. M. Soares, Phys. Rev. D {\bf 54}, 6837 (1996).
\bibitem{ISGW} N. Isgur, D. Scora, B. Grinstien and M. B. Wise, Phys. Rev. D 
{\bf 39}, 799 (1989).
\bibitem{BSW} M. Bauer, B. Stech and M. Wirbel, Z. Phys. C {\bf 34}, 103 (1987).
\bibitem{GP} E. Golowich and S. Pakvasa, Phys. Rev. D {\bf 51}, 1215 (1995).
\bibitem{IW} N. Isgur and M. B. Wise, Phys. Rev. D {\bf 42}, 2388 (1990).
\bibitem{ISGW2} D. Scora and N. Isgur, Phys. Rev. D {\bf 52}, 2783 (1995).
\bibitem{itzykson} C. Itzykson and J. B. Zuber, Quantum field theory, 
Mc-Graw-Hill, New York (1985).
\bibitem{BGM} I. Bigi, F. Gabbiani and A. Masiero, Z. Phys. C {\bf 48}, 633 
(1990).
\bibitem{BHLP} K.S. Babu, X. G. He, X. Q. Li and S. Pakvasa, Phys. Lett. B {\bf 
205}, 540 (1988). 
\bibitem{Singer} G. Eilam, A. Ioannissian. R. R. Mendel and P. Singer, Phys. 
Rev. D {\bf 53}, 3629 (1996).
\bibitem{DHT} N.G. Deshpande, X.G. He and J. Trampetic, Phys. Lett. B {\bf 367}, 
362 (1996).
\end{thebibliography}
\end{document}